# Exploring vestibulo-ocular adaptation in a closed-loop neuro-robotic experiment using STDP. A simulation study

Francisco Naveros[*], Jesús A. Garrido, Angelo Arleo, Eduardo Ros, Niceto R. Luque

*Abstract*— Studying and understanding the computational primitives of our neural system requires for a diverse and complementary set of techniques. In this work, we use the Neuro-robotic Platform (NRP) to evaluate the vestibulo ocular cerebellar adaptation (Vestibulo-ocular reflex, VOR) mediated by two STDP mechanisms located at the cerebellar molecular layer and the vestibular nuclei respectively. This simulation study adopts an experimental setup (rotatory VOR) widely used by neuroscientists to better understand the contribution of certain specific cerebellar properties (i.e. distributed STDP, neural properties, coding cerebellar topology, etc.) to r-VOR adaptation. The work proposes and describes an embodiment solution for which we endow a simulated humanoid robot (iCub) with a spiking cerebellar model by means of the NRP, and we face the humanoid to an r-VOR task. The results validate the adaptive capabilities of the spiking cerebellar model (with STDP) in a perception-action closed-loop (r-VOR) causing the simulated iCub robot to mimic a human behavior.

## I. INTRODUCTION

Computational neuroscience is evolving rapidly as regards neural simulation platforms (i.e. NEURON [1], GENESIS [2], NEST [3], EDLUT [4, 5], BRIAN [6], CARLsim [7]) and hardware spiking platforms (i.e. Spinnaker [8], Brainscales [9], TrueNorth [10]), thus covering from detailed neural cell to complex neural system modeling. Most simulation efforts in computational neuroscience are devoted to characterize neural models and/or neural system aiming at understanding the computational primitives underlying the neurophysiological substrate [11]. To this end, the cause-effect relationship between well-defined input stimulation patterns and the neural output responses is traditionally settled and studied, thereby obviating the need for a body. Nevertheless, several specialized brain regions, such as the cerebellum, have co-evolved with the body as the computational primitives in continuous interaction with the environment.

The mammalian cerebellum is pivotal in integrating the sensory information and coordinating the subsequent motor action [12], being the perfect candidate for studying its computational primitives in perception-action closed-loop setups. Experimental neuroscience uses well-established experimental setups (behavioral/cognitive tasks) which facilitate the study of the cerebellar role in motor control and adaptation. The standardization of these experimental setups helps to contrast and understand the results obtained by different research groups. Particularly, the neuroscience community widely uses the Eye blink Classical Conditioning (EBCC) [13] or/and the Vestibulo-Ocular Reflex (VOR) [14] setups in cerebellar studies. Computational neuroscience may use the outcomes from EBCC/VOR studies to further develop and validate biologically plausible cerebellar models that provide insight and offer explanation regarding the CNS computational primitives.

Replicating EBCC and VOR synthetic setups requires embodying the cerebellar network within a front-end body. Embodying not only needs for a biologically plausible cerebellar network and an actual body but also body-cerebellar efficient interfaces [15]. Early works interfacing body and mind [16] were handcrafted solutions due to the early stage of the embodiment field and the lack of specialized tools.

It was not until recent years when several initiatives aiming at building robotic platforms focused on simulating embodiment experiments appeared. Particularly, the Neurorobotics Platform (NRP) [17] was created and evolved in the framework of the Human Brain Project (HBP), a multidisciplinary project integrated by neuroscience scientists, computational neuroscience researchers, robotic engineers as well as software developers amongst others. One of the major goals of NRP lies on facilitating the implementation of simulated brain-body-environment experiments. Beyond providing the access to robotic platforms to non-experts (such as the neuroscientific community), it also facilitates the reproducibility of the results by other research groups, since not only the experiment description but also the specific experimental setup can be shared and reproduced (the source code is made available at https://github.com/EduardoRosLab/VOR_in_neurorobotics).

In this work, we present a VOR case of study using the NRP in which we show how certain cerebellar functions are supported by the underlying neural system circuitry, cell features and synaptic plasticity properties. Particularly, we depict how different plasticity mechanisms within our cerebellar model complement each other for compensating the head with the eye movement, allowing the visual stabilization of the image in the fovea.

## II. METHODS

### A. *Vestibulo-Ocular Reflex (VOR)*

The VOR is a reflexive eye movement responsible for stabilizing and centering the images on the visual field

*F. Naveros, J. A. Garrido, E. Ros and N. R. Luque are with the research Centre for Information and Communication Technologies (CITIC), Department of Computer Architecture and Technology, University of Granada, Granada, 18014, Spain (e-mail: fnaveros@ugr.es, jesusgarrido@ugr.es, eros@ugr.es, nluque@ugr.es).

A. Arleo is with the Institut de la Vision, Institut National de la Santé et de la Recherche Médicale, U968 and Centre National de la Recherche Scientifique, UMR_7210, Paris, France. He is also with the UMR_S 968, Sorbonne Universités, Université Pierre et Marie Curie Paris 06, Paris, France (e-mail: angelo.arleo@inserm.fr).

during head movements by contralateral eye movements. This reflex depends on the vestibular system which uses the semicircular canals and otolithic organs to detect rotational and translational head movements [18]. VOR does not depend on vision and it works both in light and darkness conditions.

VOR's nature is purely feed-forward since it induces prompt contralateral eye movements for compensating ipsilateral head movements. The cerebellum receives information about the head movements (signaled by the vestibular organ), the eye movements (proprioception) and the error in the fixation (signaled by the retinal slips). The cerebellar adaptation process, driven by the error signal, minimizes the retinal slips sculpting the vestibular output that is responsible for generating the compensatory motor commands, which ultimately drive the eye movement.

Our experimental setup consists of a 1 Hz rotatory VOR (r-VOR) tasks in the horizontal plane (see Fig. 1). The r-VOR task is repeated during 60 s (60 trials) so as to unveil the cerebellar adaptation process.

### B. The Neurorobotic Platform (NRP)

The NRP allows the simulation of "brain" models (neural systems implemented in NEST) connected to virtual robots able to interact with the environment (both, robot and environment implemented in Gazebo [19]). In our case of study (r-VOR task), the cerebellum plays the role of the neural structure whereas the front-end body is played by the simulated iCub robot [20].

During the embodiment simulation, the NRP closes the control loop by propagating the sensory information from the virtual iCub robot sensors to the cerebellar inputs and the motor commands from the cerebellar output to the virtual iCub robot actuators. This communication is not straightforward since our neural structure and front-end body "talk" to each other in a different language. Whilst the iCub robot uses analog signals for codifying sensory-motor information, the cerebellar model uses spikes for processing the neural information. A Rosetta stone is needed for translating information from analog to spiking signals and vice versa. This translation is done via "transfer functions" defined by the user in the NRP. These transfer functions act as neural input/output interfaces that can be interconnected via ROS [21] with different robot sensors/actuators.

### C. Cerebellar Spiking Neural Network Model

The cerebellar model (implemented in NEST) consists of six neural sub-populations (inspired in [22]): mossy fibers (MFs), granular cells (GrCs), Golgi cells (GoCs), Climbing fibers (CFs), Purkinje cells (PCs) and Vestibular Nuclei (VN) cells representing the flocculus [23] (a small lobe of the cerebellum essential for the VOR).

The MFs convey the sensory information from the vestibular organ, providing the sensory inputs to the cerebellar network. These MFs project excitatory afferents onto GrCs, GoCs and VN cells. The GoCs, in turn, inhibit the GrCs. The CFs project excitatory afferents onto PCs and drive the teaching signal responsible for cerebellar adaptation. The PCs integrate and correlate the sensory activity from parallel fibers (PFs) (i.e. the axons of GrCs) with the error-driven activity coming from the CFs. The synaptic weights at PFs-PCs connections are therefore adapted, causing the PC firing output to be adjusted. Finally, PCs inhibit VN cells, thus closing the cerebellar loop. The VN cells integrate and correlate the sensory activity coming from MFs with the inhibitory activity coming from PCs. The synaptic weights at MFs-VN cells are therefore adapted, causing the VN output to be sculpted. VN cells generate the cerebellar output activity that ultimately excites the oculomotor neurons and drives the eye movement. This neural structure is described in table I and Fig. 2.

**Mossy fibers (MFs):** This layer consists of 100 fibers that are connected to 20 GrCs, 1 GoCs and 200 VN cells each (excitatory connections).

**Golgi cells (GoCs):** This layer consists of 100 Leaky Integrate-and-Fire (LIF) neurons that are connected to 80 GrCs each (inhibitory connections).

**Granular cells (GrCs):** This layer consists of 2000 LIF neurons that are connected to 200 PCs each (excitatory connections). GrCs transform the sensory activity coming from the MFs into somatosensory neural activity by generating spatiotemporal patterns that are correlated downstream with the error activity at PC layer.

**Climbing fibers (CFs):** This layer consists of 200 fibers that are connected to one PC each. Each CF activation generates a large event (strong excitatory synapse) in the PC soma that is correlated with the somatosensory activity coming through GrCs axons.

**Purkinje cells (PCs):** This layer consists of 200 LIF neurons that are connected to one VN cell each (inhibitory connections). This layer is divided into two groups of 100

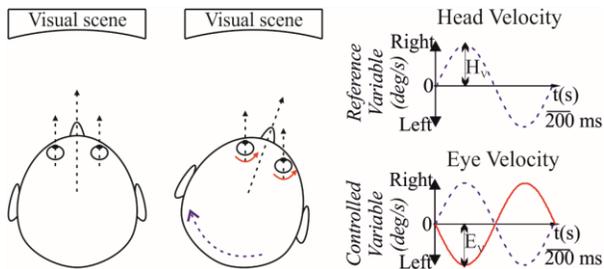

Figure 1. Vestibulo-ocular reflex (VOR) experiment. VOR stabilizes the visual field during horizontal head rotation tests (r-VOR) by producing contralateral eye movements.

TABLE I. NEURAL NETWORK TOPOLOGY

| Neurons | | Synapses | | | |
|---|---|---|---|---|---|
| Pre-synaptic | Post-synaptic | Number | Type | Initial weight | Weight range |
| 100 MFs | 2000 GrCs | 40000 | Exc | [0.00048 0.00072] | - |
| 100 MFs | 100 GoCs | 100 | Exc | 0.0006 | - |
| 100 MFs | 200 VN | 20000 | Exc | 0.001 | [0.0005 0.0070] |
| 100 GoCs | 2000 GrCs | 160000 | Inh | [0.00016 0.00024] | - |
| 2000 GrCs | 200 PCs | 400000 | Exc | 0.005 | [0.000 0.010] |
| 200 CFs | 200 PCs | 200 | Exc | 0.005 | - |
| 200 PCs | 200 VN | 200 | Inh | 0.00002 | - |

cells each that are responsible for shaping the VN agonist /antagonist output.

**Vestibular Nuclei (VN) cells:** This layer consists of 200 LIF neurons that generate the cerebellar output. This layer is divided into two groups of 100 cells each that are responsible for driving the activation of the agonist/antagonist pair of the eye muscles.

The input fibers (MFs and CFs) are modeled in NEST as "parrot_neurons" since they merely propagate the sensory and error activity to the target neurons whereas the GrCs, GoCs, PCs and VN neurons are modelled as "IF_cond_alpha" neurons (a standard Leaky Integrate-and-Fire model).

### D. Plasticity Mechanisms

The synaptic weight distribution responsible for the cerebellar input-output response has made adaptive through STDP mechanisms at two different levels. An in depth review about these plasticity mechanisms can be found in [22]. These STDP mechanisms were taken from EDLUT and re-implemented and adapted for NEST.

**PFs–PCs synaptic plasticity:** The long-term depression (LTD) / long-term potentiation (LTP) balance at PFs–PCs synapses is based on Eqs. 1-2:

$$LTD\ \Delta w_{PF_j-PC_i}(t) = \int_{-\infty}^{CF_{spike}} k\left(\frac{t-t_{CF_{spike}}}{\tau_{LTD}}\right) \cdot \delta_{PF_{spike}}(t) \cdot dt \quad (1)$$

$$LTP\ \Delta w_{PF_j-PC_i}(t) = \propto \cdot \delta_{PF_{spike}}(t) \quad (2)$$

Where $\Delta W_{PFj-PCi}(t)$ denotes the weight change between the $j^{th}$ PF and the target $i^{th}$ PC; $\tau_{LTD}$ is the time constant that compensates the sensorimotor delay; $\delta_{GC}$ is the delta Dirac function corresponding to an afferent spike from a PF; and the kernel function $k(x)$ is defined as:

$$k(x) = e^{-x} \cdot \sin(x)^{20} \quad (3)$$

The effect on the presynaptic spikes arriving through PFs is maximal over the 100 ms time window before CF spike arrival, thus accounting for the sensorimotor pathway delay [24-27]. The amount of LTP at PFs is fixed (Eq. 2), with an increase in synaptic efficacy equal to α each time a spike arrives through a PF to the targeted PC.

**MFs–VN synaptic plasticity:** The LTD/LTP dynamics at MFs – VN synapses is based on Eqs. 4-5:

$$LTD\ \Delta w_{MF_j-VN_i}(t) = \int_{-\infty}^{\infty} k\left(\frac{t-t_{PC_{spike}}}{\sigma_{MF-VN}}\right) \cdot \delta_{MF_{spike}}(t) \cdot dt \quad (4)$$

$$LTP\ \Delta w_{MF_j-VN_i}(t) = \propto \cdot \delta_{MF_{spike}}(t) \quad (5)$$

with $\Delta W_{MFj-VNi(t)}$ denoting the weight change between the $j^{th}$ MF and the target $i^{th}$ VN. $\sigma_{MF-VN}$ standing for the temporal width of the kernel; $\delta_{MF}$ representing the delta Dirac function that defines a MF spike; and the integrative kernel function $k(x)$ defined as:

$$k(x) = e^{-|x|} \cdot \cos(x)^2 \quad (6)$$

The STDP rule defined by Eq. 4 produces a synaptic efficacy decrease (LTD) in the MF weight when a spike from the PC reaches the targeted VN neuron. The amount of synaptic decrement (LTD) depends on the activity arrived through the MFs. This activity is convolved with the integrative kernel defined in Eq. (6). This LTD mechanism considers those MF spikes that arrive after/before the PC spike arrival within the time window defined by the kernel. The amount of LTP at MF - VN synapses is fixed, with an increase in synaptic efficacy equal to α each time a spike arrives through a MF to the targeted VN.

### E. iCub robot

The humanoid iCub robot, used as front-end body in the r-VOR task, can sense its own body position (proprioception) and movement (using accelerometers and gyroscopes) [20]. These sensors are used to emulate the vestibular and proprioceptive signals.

Performing the r-VOR task requires moving the iCub head (controlled by the neck) and eyes in the horizontal plane. The motors articulating the eye and head movements are controlled by using either position or velocity commands. Since the cerebellar r-VOR output drives the eye velocity movements, the iCub motors are controlled using velocity commands.

### F. Closing the control-loop; the transfer functions

The sensory-motor information is to flow between the humanoid iCub robot (simulated in Gazebo) and the cerebellar neural network (simulated in NEST). This communication requires both the periodic synchronization and the propagation of the sensory-motor information between simulators (NEST & Gazebo). The control cycle is closed through NRP allowing the synchronization of both simulators each 10 ms, whereas the NRP transfer functions allow propagating the sensory-motor information. We implemented three transfer functions accounting for:

a) The sensory information propagated from the neck iCub sensors to the cerebellar MFs.

b) The error-related sensory information propagated from iCub sensors to the cerebellar CFs.

c) The neural motor commands propagated from the cerebellar output neurons (VN) to the iCub actuators.

These three transfer functions also account for translating the sensory-motor information from analog to spiking signals and vice versa. The fourth transfer function is dedicated to operate the head. Fig. 2 shows a complete diagram of the control loop implemented for the r-VOR task, including the cerebellar model, the iCub robot and the four transfer functions aforementioned.

**Head twist**: This transfer function generates and propagates each 10 ms a motor commands that produces the sinusoidal head movement that trigger the r-VOR cerebellar adaptation.

**Sensory activity**: This transfer function reads each 10 ms the iCub head encoders (analog signals codifying the head position and velocity) and translates their analog information into a neural population coding representation for the MF layer. 100 Poisson distribution processes, whose mean depends on the head position/velocity, generate the activity of each MF. The transfer function incorporates a set of receptive fields (Gaussian like curves) that covers all the possible head position/velocity values. Depending on the head position/velocity, the receptive field of each MF is activated to a greater or lesser degree. Each receptive field activation value set the mean firing rate for each Poisson distribution process. Consequently, each fiber becomes sensitive to a specific range of positions or velocities.

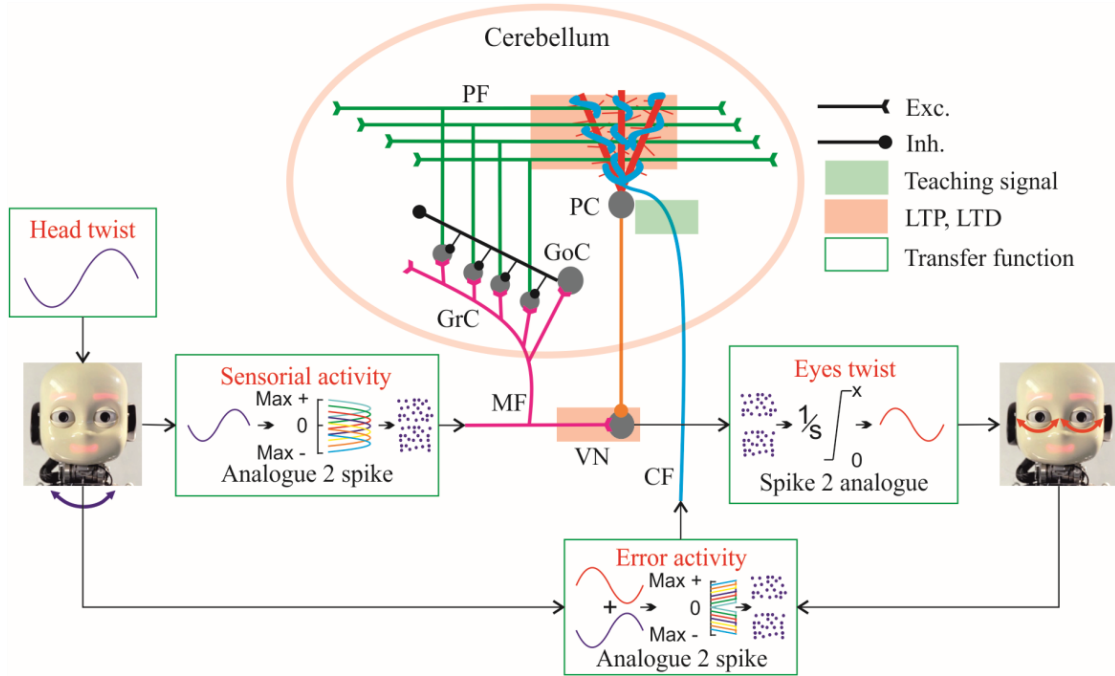

Figure 2. Cerebellar control loop for r-VOR task using an iCub robot as front-end body. The head twist transfer function generates the motor commands that produces the head movement and trigger the r-VOR response. The sensory and error activity transfer functions read the sensory-motor signals from the iCub encoders and propagate this information through MFs and CFs respectively after an analog-to-spike conversion. The cerebellar model processes this input activity and generates the corresponding output response through the VN. Finally, the VN output spike activity is transformed into an analog signal that controls the eye movement. (MF: Mossy Fiber, GrC: Granular Cell, GoC: Golgi Cell, PF: Parallel Fiber, PC: Purkinje Cell, CF: Climbing Fiber, VN: vestibular nuclei).

**Eyes twist**: This transfer function calculates the eye motor commands integrating the agonist and antagonist VN activity each 10 ms. These motor commands are then sent to the eye actuator causing the eye movement.

**Error activity:** This transfer function reads the sensory information from the head and eye movement encoders and computes the retinal slip. This retinal slip is translated into a neural population coding representation at the CF layer. 200 Poisson distribution processes whose mean depends on the intensity of the error sensed (1 Hz for minimal error, 10 Hz for maximal error), generate the activity of each CF. This transfer function incorporates a set of receptive fields (sigmoid like curves) that covers the error range (all the possible retinal slip values). Depending on the sensed error, the sigmoidal receptive field of each CF is activated to a greater or lesser degree. Each receptive field activation value set the mean firing rate for each Poisson distribution process making each CF sensitive to a specific error range.

### III. RESULTS

We use a robotic r-VOR task as case of study for our embodiment scenario. The cerebellar adaptation process is performed in the horizontal plane using a cerebellar spiking model as neural structure, an iCub robot as front-end body and the NRP orchestrating the body-mind communication.

The cerebellar adaptation process lasts over 60 s, 1 s per trial (1 Hz sinusoidal head movement). We start out with a blank sheet, the untrained cerebellar structure does not provide for any head movement compensation. The initial synaptic weights at the plastic sites (PFs-PCs and MFs-VN) are to be adapted. The starting error is maximal since cerebellar adaptation is not deployed yet (see Fig 3.A) and the motor output is negligible (see Fig 3.B left panel). After 20 s the cerebellar adaptation process starts providing certain eye compensation for the head movement (see Fig 3.A and Fig 3.B central panel). At the end of the learning process (60 s), the error converges towards zero indicating an almost perfect eye compensation of the head velocity (see Fig 3.A and Fig 3.B right panel).

The cerebellar adaptation process shapes the spiking activity response of the neurons within the CF-PC-VN cerebellar sub-circuitry. MFs and PFs (arising from GrCs) supply for the baseline neural activity that VN cells and PCs need for their neural operation (see Fig. 3.D). The STDP mechanisms are to adjust the synaptic efficacy of the afferent nerve fibers at MFs–VN and PFs-PCs and, consequently, to modulate the VN and PC neural activity. The beginning of adaptation process marks the maximal CF activation frequency. The starting error is maximal as the CF activity that drives learning. The two antagonist micro-complexes in which CFs are divided (sensing the error obtained in clockwise and anticlockwise direction) are spiking proportionally to the sensed error at 10 Hz on average. The STDP mechanism at the cerebellar molecular layer cannot silence the PC output yet (decreasing the PFs-PCs synaptic efficacy when needed); the PC inhibition activity is maximal so as the VN firing output is minimal (see Fig 3.C, left panel). After 20 s, the adaptation process is halfway; The CFs firing rate is diminishing as the cerebellar output begins to stabilize. The STDP at the molecular layer starts blocking PC output when needed. The STDP at VN afferents uses these PC periods of silence to increase the synaptic efficacy of the afferent nerve fibers at MFs–VN causing the VN spiking output starts the eye compensation for the head movement (see Fig 3.C, middle panel). Eventually, after 60 s, the adaptation process converges; CFs firing rate reaches its minimum (1-2 Hz), both STDPs mechanism are able to regulate the blockage of PC and VN output activity when needed and the VN spiking output completely deploys the eye compensation for the head movement (see Fig 3.B and C, right panels).

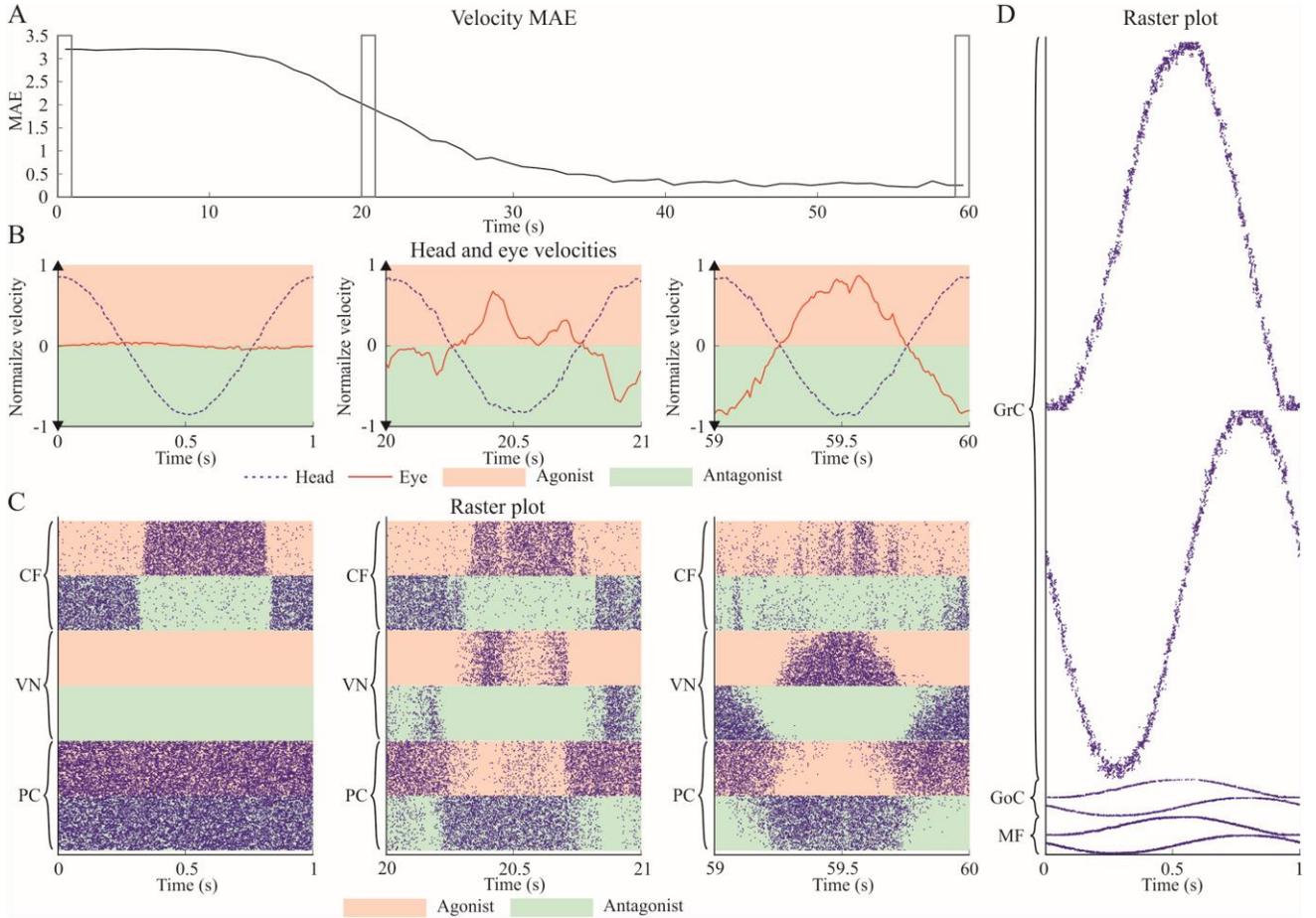

Figure 3. iCub robot signals and cerebellar neural activity in a r-VOR task. A) Plot depicting the Mean Absolute Error (MAE) between head and eye velocities that are obtained for each trial during the r-VOR cerebellar adaptation process. For a perfect cerebellar head velocity compensation, the eye velocity signal must be in counter-phase with equal amplitude. B) Head and eye velocity trajectories at three different learning stages; initial, middle and final cerebellar adaptation process. C) CF, VN and PC neural activity at three different learning stages: initial, middle and final cerebellar adaptation process. The neural activity of these layers evolves with the cerebellar adaptation mechanisms. The CF-PC-VN cerebellar sub-circuitry is divided in two antagonist micro-complexes responsible for head velocity compensation in clockwise and anticlockwise direction. D) MF, GoC and GrC neural activity. The vestibular signals produced as consequence of the head velocity remain unchanged as well as the neural activity caused by these vestibular signals.

The consequence of the complete deploy of the r-VOR cerebellar adaptation is a stabilization of the image in the retina during the head rotation by a compensatory ocular movement in the opposite direction with the same amplitude, thus preserving the image within the center of the visual field. We aim to illustrate this in Fig. 4. NRP facilitates the visual verification of the correct adaptation of the reflex. Fig. 4 left panel depicts an iCub robot that is not capable of performing the eye compensation. The eyes move conjointly with the head; the visual field is not centered. Conversely, Fig. 4 right panel depicts an iCub robot after cerebellar adaptation that is able to compensate the head with the eyes movement; the visual field is stabilized and centered.

IV. CONCLUSIONS

In this work, we present one of the first cerebellar embodiment case-of-study that builds bridges between the neural cerebellar structure (its inherent plasticity mechanisms, neuron model, sub-circuitry and neural coding) and the behavioral outcomes observed in biology (i.e. r-VOR). Our spiking cerebellar model is able to effectively learn the vestibule ocular reflex thanks to the two STDP adaptive mechanisms located at the cerebellar molecular layer and at VN afferents coming from MFs. These two STDP mechanisms operate conjointly to shape the cerebellar neural activity that ultimately generate the eye motor commands that compensate the head movement in the iCub robot. The STDP mechanisms adapt the synaptic weights of PFs-PCs and MFs-VN afferents according to the teaching signal coming from CFs sensing the error and the sensorial information sensed by the vestibular system through MFs.

The solution here proposed hypothesizes on the operation of the cerebellar computational primitives from a bottom-up perspective; starting at a neuronal level, through network and system level, and ending at behavioral level. Accessing the entire perception-action loop helps us to discuss the whys and wherefores of what is going on with the cerebellar neural structure, the front-end body (iCub) and the body-mind dialogue. The main aim here is to ***provide an embodiment case-of-study*** as a basis for drawing humanoid-human analogies that may drive basic cerebellar research by proposing working cerebellar hypotheses that can be either refuted or validated from a cellular or neural network point of view.

ACKNOWLEDGMENT

This research was supported by the "contrato puente" UGR fellowship (F. Naveros), the Juan de la Cierva Spanish fellowship (N. R. Luque), a grant from the European Union (Cerebsensing, H2020-653019 to J. A. Garrido), the European Union ER (HBP-SGA1, H2020-RIA. 720270; HBP-SGA2, H2020-RIA. 785907) and the National Grant (TIN2016-81041-R partially funded by FEDER).